\documentclass[prb,twocolumn,floatfix,showpacs,preprintnumbers, 
               superscriptaddress,amsmath,amssymb]{revtex4}
\usepackage[dvips]{graphics}
\usepackage{psfrag}
\usepackage{graphicx}
\usepackage{epsf}
\usepackage{umlaut}

\def\iv{{$I-V$ characteristics$\;$}}

\begin{document}
\title{Super-poissonian noise, negative differential conductance, and
relaxation effects in transport through molecules, quantum dots and nanotubes}

\author {Axel Thielmann} 
\affiliation{Forschungszentrum Karlsruhe, Institut f\"ur Nanotechnologie,
76021 Karlsruhe, Germany}

\author {Matthias H. Hettler} 
\affiliation{Forschungszentrum Karlsruhe, Institut f\"ur Nanotechnologie,
76021 Karlsruhe, Germany}

\author {J\"urgen K\"onig} 
\affiliation{Institut f\"ur Theoretische Physik III, Ruhr-Universit\"at
             Bochum,  44780 Bochum, Germany}

\author {Gerd Sch\"on} 
\affiliation{Forschungszentrum Karlsruhe, Institut f\"ur Nanotechnologie,
76021 Karlsruhe, Germany}
\affiliation{Institut f\"ur Theoretische Festk\"orperphysik,
Universit\"at Karlsruhe, 76128 Karlsruhe, Germany}

\date{\today}

%%%%%%%%%%%%%%%%%%%%%%%%%%%%%%%%%%%%%%%%%%%%%%%%%%%%%%%%%%%%%%%%%%%%%%%%%%%%%%
\begin{abstract}
We consider charge transport through a nanoscopic object, e.g.\ single molecules, short
nanotubes, or quantum dots, that is weakly coupled to metallic
electrodes. We account for several levels of the molecule/quantum dot with
level-dependent coupling strengths, and allow for relaxation of the
excited states. The current-voltage characteristics as well as the
current noise are calculated within first-order perturbation  
expansion in the coupling strengths.
For the case of asymmetric coupling to the leads we predict
negative-differential-conductance  accompanied with super-poissonian 
noise. Both effects are destroyed by fast relaxation processes.
The non-monotonic behavior of the shot noise as a function of bias and relaxation rate
reflects the details of the electronic structure and level-dependent coupling strengths.
\end{abstract}

\maketitle
%%%%%%%%%%%%%%%%%%%%%%%%%%%%%%%%%%%%%%%%%%%%%%%%%%%%%%%%%%%%%%%%%%%%%%%%%%%%%%

\section{Introduction} 
The field of molecular electronics\cite{review} is driven by the quest
for functional electronic devices that are smaller than those produced
by standard semiconductor technology.
The microscopic size and the reproducibility in the production of
molecules provide decisive advantages even if many, identical,
molecules should be needed to build fault-tolerant devices
A negative differential conductance (NDC), a promising feature for functional 
devices, has been found recently in organic molecules.\cite{chen-etal}
Furthermore,  many-body effects such as the Coulomb blockade and  
Kondo effect, known from semiconductor quantum 
dots,\cite{goldhaber}
have been observed in molecular devices\cite{low_temp,cocomplex}
and carbon nanotubes.\cite{lindelof,schoenenberger,vanderZant} 

Since the typical single-particle level spacing of quantum dots 
(or short nanotubes) is small -- often only a fraction of
a meV -- low temperatures are required for the  
resolution of transport through individual levels.
Low temperatures are also helpful for the observation of quantum or shot noise.
For example, at temperatures above 30K noise transport through
molecules between gold break  junctions\cite{reichert_thesis} appears
to be dominated by 1/f-like noise, believed to be generated by 
thermally-induced fluctuations of the gold atoms.
Such effects are suppressed at sub-Kelvin temperatures, at which the shot 
noise associated with the discreteness of the charge of the 
transfered electrons\cite{blanter} can be detected.
Both the current and the shot noise depend on details of the discrete level 
spectrum and the coupling strengths of these levels to the 
electrodes.\cite{haug,thielmann,kiesslich}
The combined measurements of current and shot noise, thus, provide a
'spectroscopic' tool to gain information about the level structure.
%and the couplings of the levels to the electrodes.

Negative differential conductance through multi-level systems can occur when 
two adjacent levels have different coupling strengths to the leads.
Once the level with weaker coupling is occupied, transport through the
other level is suppressed, which reduces the total current. 
For molecules, it is well known that the coupling of different molecular
orbitals to the electrodes may vary strongly due to differences in the
spatial structure of the corresponding wave functions.\cite{hsw}  
In  metallic single-walled nanotubes (SWNTs), two bands cross
the Fermi surface as the doping level is varied. 
For short tubes, these bands break up in a set of single-particle 
levels, separated by a level spacing $\delta E$ of about 1 meV or less 
(see also Ref.~\onlinecite{schoenenberger} for multi-walled nanotubes).
In general, these levels  differ in their spatial structure and
coupling strength, particularly if  they derive from two different
bands at different points of the one-dimensional Brillouin zone. 
For semiconducting nanotubes similar considerations hold.\cite{vanderZant}
NDC has also been observed in semiconductor quantum dots.\cite{ndc_qdots}

To study transport through systems which display a NDC, we start from
an effective  
model of a few single-particle levels with couplings to the electrodes
which vary strongly from level to level, and which also may 
differ for the source and drain electrodes.
Furthermore, we include the possibility of relaxation among the levels:
at finite bias voltage, electrons might enter the molecule (quantum
dot or nanotube) at a high-lying excited state. 
Provided that the relaxation is fast as compared to the tunneling, the 
molecule might relax to the ground state or other low-lying state 
before the electron has the chance to leave the molecule. 
The relaxation is accompanied by the emission of a boson, either a 
photon or a phonon.
Such relaxation processes can have a strong impact on the negative
differential conductance by destroying the blocking mechanism.\cite{hsw}
Related models (without coupling to a bosonic bath and relaxation) were 
studied in Refs.~\onlinecite{bulka,nazarov,kiesslich2}.

The main purpose of this work is to study shot noise for the model
described above in the regime where NDC might occur.
We predict that in the absence of relaxation, the NDC is accompanied with
super-poissonian noise. 
This is formally similar to transport through semi-conducting
resonant tunneling devices,\cite{RTD} though the origin of NDC in these devices
(chemical potential passes through the semiconductor band edge)
is entirely different from the one discussed here.
Relaxation processes enhance the current and reduce the noise in the NDC
regime.
The shot noise shows rather rich behavior depending on the coupling and 
relaxation strength.
In particular, we find that the shot noise is a non-monotonic function of
the relaxation rate.
This behavior contrasts with the current, which monotonically increases
as the relaxation rate becomes larger.
We are able to present analytic results for the shot noise, which might be 
useful for the interpretation of future experiments.
We also relate our results to  models of transport through several quantum dots.

\section{The model}
As a model for electron transport through a molecule or nanotube  with $M$ 
molecular orbitals (levels) and Coulomb interaction we consider a generalized
Anderson impurity model coupled to a bosonic bath, described by the 
Hamiltonian
$\hat H = \hat H_{\rm L} + \hat H_{\rm R} + \hat H_{\rm M} + 
 \hat H_{\rm T,L} + \hat H_{\rm T,R} + \hat H_{\rm ph}$
with: \\
\begin{eqnarray}
&& \hat H_{r} = \sum_{k \sigma}\epsilon_{k \sigma r} a_{k \sigma r}^{\dag}
            a_{k \sigma r},  \\
&&\hat H_{\rm M} = \sum_{l \sigma}\epsilon_{l \sigma} c_{l \sigma}^{\dag}c_{l \sigma} 
 + U\sum_l n_{l\uparrow}n_{l\downarrow}  + E_c (\sum_{l\sigma}{n_{l\sigma}})^2 \\ 
%            \frac{\Delta_{ex}}{2}\sum_{\sigma,\sigma',l\neq l'}  
%            c_{l \sigma}^{\dag} 
%            c_{l' \sigma'}^{\dag}c_{l \sigma'}c_{l' \sigma}, \\
&&\hat H_{{\rm T},r}= \sum_{l k \sigma}(t^r_l a_{k \sigma
            r}^{\dag} c_{l \sigma} 
            + h.c.),  \\   
&&\hat H_{\rm ph}= \sum_{q}\omega_{q}
            d_{q}^{\dag} d_{q}+\sum_{q,\sigma,l,l'} 
            g^{l,l'}_{q} (d_{q}^{\dag}+ d_{q}) 
            c_{l \sigma}^{\dag}c_{l' \sigma},
\end{eqnarray}
where $l=1 \ldots M$ and $r={\rm L},{\rm R}$.
Here, $\hat H_{\rm L}$ and $\hat H_{\rm R}$ model the non-interacting
electrons with density of states 
$\rho_e= \sum_k \delta(\omega -\epsilon_{kr})$ 
in the left and right electrode ($a_{k \sigma r}^{\dag}, a_{k \sigma r}$ are the Fermi
operators for the states in the electrodes). The molecule term
$\hat H_{\rm M}$ describes a 
"molecule" with $M$ relevant molecular orbitals of energy 
$\epsilon_{l \sigma}$ and Coulomb interaction on the molecule
($c_{l\sigma}^{\dag}, c_{l\sigma}$ are Fermi operators for the molecular 
levels, and $n_{l\sigma}=c_{l \sigma}^{\dag}c_{l \sigma}$ is the number 
operator). 
The charging energy $E_c$ accounts for the classical
energy cost to add charge on a confined system with many electrons and ions that are not explicitly considered in the Hamiltonian. In addition, the 
Hubbard-like term with energy $U$ punishes double occupancy within the same orbital.
These two kinds of interaction terms are the most important parts of the full two-body
interactions present in a real molecule. Other terms could be considered by much more
elaborate models, as done in Ref.~\onlinecite{hettler_prl} for computation of the \iv .
For the NDC/relaxation effects on the shot noise that we wish to study, the above simple molecule model suffices.
Tunneling between leads and molecule levels is modeled by 
$\hat H_{{\rm T},{\rm L}}$ and $\hat H_{{\rm T},{\rm R}}$.
The coupling strength is characterized by the intrinsic line width 
$\Gamma^r_l = 2 \pi |t^r_l|^2 \rho_e$, where $t^r_l$ are the
tunneling matrix elements. 
In order to allow for relaxation between different molecular levels, we add
$\hat H_{\rm ph}$, which describes a bosonic bath (where $d_{q}^{\dag}, d_{q}$ 
are the corresponding Bose operators) coupled to the
molecule by the coupling constants $g^{l,l'}_{q}, l\neq l'$.
This allows relaxation processes where electrons on the molecule can change 
the orbital by emitting or absorbing a boson.
Note that a diagonal coupling, $l=l'$, would not be associated with relaxation 
but would give rise to ``boson-assisted tunneling'' leading to additional 
steps in the \iv when the boson bath has a 
discrete spectrum.\cite{boese_epl,aleiner,diagrams}  
Since in the present paper we are not interested in those boson-assisted 
tunneling processes, we take into account off-diagonal coupling contributions
only.
To be specific, we assume in the following that the bosonic bath consists of
photons, although vibrational effects due to phonons could also be described 
within our model.
For simplicity, we assume the constants $g^{l,l'}_{q} = (1-\delta_{l,l'}) 
g_{\rm ph}$ to be independent of $l,l'$ and $q$, and introduce the 
coupling $\alpha_{\rm ph}$ as 
$\alpha_{\rm ph} (\omega)=  2 \pi g^2_{\rm ph} \rho_{\rm b}(\omega)$, where $\rho_{\rm b}(\omega)= \sum_q \delta(\omega -\omega_q)$ is the density of states of the bosonic bath
For the relaxation due to photons we choose a power law behavior  $\rho_{\rm b}(\omega) \propto \omega^3$,
corresponding to photons with $3$ spatial degrees of freedom. 
For the case of phonon-mediated relaxation, which we are not going to discuss
in detail in the present paper, the density of states is sharply peaked around 
the vibration modes of the molecule.
%The exponent $s$ depends on the 
%underlying microscopic realization of the boson bath 
%(in our case for the
%description of photons with 3 spatial degrees of freedom we chose $s=3$) 
%and $\omega$ is the energy to be integrated over for our later calculations
%of the transition rates. 
%Again $\alpha_{\rm ph}$ can be related to $\Gamma$ via
%a dimensionless coupling parameter. 
The above model is an extension of the Anderson impurity model with
one level (and in the absence of relaxation effects), which was described
and discussed in Ref.~\onlinecite{thielmann}. 

We are interested in transport through the molecule, in particular in the
current $I$ and the (zero-frequency) current noise $S$.
They are related to the current operator 
$\hat I = (\hat I_{R} - \hat I_{\rm L})/2$, with $\hat{I}_r = -i(e/\hbar) 
\sum_{l k \sigma} \left( t^r_l a_{k \sigma r}^{\dag} c_{l\sigma} - h.c.\right)$ 
being the current operator for electrons tunneling into lead $r$, by 
$I = \langle \hat{I} \rangle$ and 
\begin{equation}
S = \int_{-\infty}^{\infty} dt \langle \delta \hat{I}(t) \delta \hat I(0) 
+ \delta \hat{I}(0) \delta \hat I(t) \rangle
\end{equation}
where 
$\delta \hat I(t)=\hat I(t)-\langle \hat I \rangle$.

\section{Diagrammatic Technique}
For the calculation of the current $I$ and current noise $S$, we use the
diagrammatic technique developed in Ref.~\onlinecite{diagrams}
and expanded for the description of the noise in
Ref.~\onlinecite{thielmann}.
In lowest-order perturbation theory in the coupling strengths $\Gamma$, the 
following expressions the current and the noise were found:
\begin{equation}
  I = {e\over 2\hbar} {\bf e}^T {\bf W}^{I}
  {\bf p}^{{\rm st}}
\label{I1}
\end{equation}
\begin{equation}
  S = {e^2\over \hbar} {\bf e}^T \left( 
    {\bf W}^{II} {\bf p}^{{\rm st}} + {\bf W}^{I} 
    {\bf P} {\bf W}^{I} {\bf p}^{{\rm st}} \right).
\label{S1}
\end{equation}
The bold face indicates matrix notation related to the molecular state
labels $\chi$ (for the $M$ level system there are $4^M$
different states). The vector $\bf e$ is given by $e_\chi = 1$ for 
all $\chi$. 
The zeroth-order stationary probabilities ${\bf p}^{{\rm st}}$  can be expressed in terms of first-order
transition rates $W_{\chi,\chi'}$ (forming a matrix $\bf W$) between two
molecular states $\chi$ and $\chi'$ as:
\begin{equation}
  {\bf p}^{{\rm st}}=  (\tilde {\bf W})^{-1} {\bf v}.
\end{equation}
The matrix $\bf \tilde W$ is identical to $\bf W$ but with 
one (arbitrarily chosen) row $\chi_0$ being replaced with 
$(\Gamma,...,\Gamma)$. Then the vector $\bf v$
is given by $v_\chi=\Gamma \delta_{\chi \chi_0}$.
The total transition rates $W_{\chi,\chi'}$ (in the absence of relaxation) 
are the sum of transition rates
associated with electron tunneling through either the left or the right 
barrier, $W_{\chi,\chi'} = W_{\chi,\chi'}^{\rm R} + W_{\chi,\chi'}^{\rm L}$.

The matrix elements of ${\bf W}^{I}$ and ${\bf W}^{II}$ are given by
$W_{\chi,\chi'}^{I}=(W_{\chi,\chi'}^{\rm R}-W_{\chi,\chi'}^{\rm L})
(\Theta(N_{\chi}-N_{\chi'})-\Theta(N_{\chi'}-N_{\chi}))$ and
$W_{\chi,\chi'}^{II}= {1 \over 4} (W_{\chi,\chi'}^{\rm R}+W_{\chi,\chi'}^{\rm L})
(1-2~\delta_{\chi \chi'})$, where $N_{\chi}$ is the total number of electrons 
on the molecule within the state $\chi$.
The indices $I$ or $II$ indicate that one or two vertices in the corresponding
diagram are due to current operators present in the definition for the
current $I$ and the noise $S$.

The matrix ${\bf P}$ is associated with the propagation between two blocks
${\bf W}^{I}$ containing one current operator each.
To lowest order in $\Gamma$, 
\begin{equation}
  {\bf P}=  (\tilde {\bf W})^{-1} {\bf Q}
  \label{def_P}
\end{equation}
with $Q_{\chi' \chi} = (p_{\chi'}^{\rm st}-\delta_{\chi', \chi}) 
(1-\delta_{\chi',\chi_0})$, i.e., ${\bf P}$ is of order $\Gamma^{-1}$, 
thus leading to a non-vanishing contribution of the second
part in Eq.~(\ref{S1}) even in lowest (first) order perturbation
theory in the coupling to the electrodes. 
Similar expressions for the calculation of current and noise were derived by 
other means by Hershfield {\it et al.} 
in Ref.~\onlinecite{hyldgaard} and Korotkov in Ref.~\onlinecite{korotkov}.\\

In order to include relaxation processes, we need to extend the theory
of Ref.~\onlinecite{thielmann} by introducing corresponding transition rates
$W_{\chi,\chi'}^{\rm ph}$.
Assuming weak coupling to the bosonic bath (in addition to weak tunneling), we 
only keep contributions to either first order in $\alpha_{\rm ph}$ or to
first order in $\Gamma$.
The total transition rates are, thus, given by $W_{\chi,\chi'} = 
W_{\chi,\chi'}^{\rm L}+W_{\chi,\chi'}^{\rm R} + W_{\chi,\chi'}^{\rm ph}$, where
$W_{\chi,\chi'}^{\rm ph}$ describe pure relaxation while 
$W_{\chi,\chi'}^{\rm L}$ and $W_{\chi,\chi'}^{\rm R}$ models pure tunneling.
The tunneling rates are given by 
\begin{eqnarray}
& W_{\chi',\chi}^{r}= 2\pi \sum_{l,\sigma} & 
   \left[ \gamma_{rl\sigma}^{+}(E_{\chi',\chi})
   | \langle \chi'| c_{l\sigma}^{\dag} | \chi \rangle |^2 \right. \nonumber \\ 
& \,  &  + \left. \gamma_{rl\sigma}^{-}(-E_{\chi',\chi})
   |\langle \chi' | c_{l\sigma} | \chi \rangle |^2 \right] 
\end{eqnarray}
for $\chi' \neq \chi$, together with $W_{\chi,\chi}^{r} = 
- \sum_{\chi' \neq \chi} W_{\chi',\chi}^{r}$, where
$\gamma_{rl\sigma}^{\pm}(E_{\chi',\chi})=\Gamma_{r}^l/2\pi~f^{\pm}
(E_{\chi',\chi} - \mu_r)$. 
%*** define $\gamma$ ***
The bosonic rates are
\begin{equation}
 W_{\chi',\chi}^{\rm ph}= 2 \pi \sum_{l \neq \bar l}  
   b(E_{\chi',\chi})
   |\langle \chi' | c_{l\sigma}^{\dag}c_{\bar l \sigma} | \chi \rangle |^2 .
\end{equation}
for $\chi' \neq \chi$, and $W_{\chi,\chi}^{\rm ph} = 
- \sum_{\chi' \neq \chi} W_{\chi',\chi}^{\rm ph}$,
with $b(E_{\chi',\chi})= \alpha_{\rm ph}(E_{\chi',\chi})/2\pi~n_{\rm b}(E_{\chi',\chi})$.
Here,  $f(x)=f^{+}(x)=1-f^{-}(x),~n_{\rm b}(x)=n_{\rm b}^{+}(x)=n_{\rm b}^{-}(-x)$ are the
Fermi and Bose functions and
$E_{\chi',\chi}= E_{\chi'}-E_{\chi}$ is the energy difference
between the many-body states $ \chi$ and $\chi'$.
While the presence of relaxation leads to a modification of
${\bf p}^{{\rm st}}$ and ${\bf P}$, the matrices ${\bf W}^{I}$ and 
${\bf W}^{II}$ are not affected.

The rules for calculating the irreducible blocks ${\bf W}$ describing electron
tunneling and relaxation are as follows:

1) For a given order $k$ draw all topologically different diagrams with
$2k$ vertices connected by $k$ tunneling (electron) lines or boson (photon)
lines (for orders $k\ge 2$ both kinds of lines might be contained in a diagram).
Assign the energies $E_{\chi}$ to the propagators, and
energies $\omega_l$ ($l=1,...,k$) to each one of these lines.

2) For each of the ($2k-1$) segments enclosed by two adjacent vertices
there is a resolvent $1/(\Delta E_j+i0^+)$ with $j=1,...,2k-1$, where 
$\Delta E_j$ is the difference of the left-going minus the right-going energies.

3) Each vertex containing dot operators $B_n$ (with $n$ different operator
   structures) gives rise to a matrix element
%   $B^n_{\chi',\chi}=
   $\langle \chi' | B_n | \chi \rangle$, where $\chi\,(\chi')$ 
   is the dot state 
   entering (leaving) the vertex with respect to the Keldysh contour
   (for our model we have: $B_1=c_{l\sigma}^{\dag}, B_2=c_{l\sigma}, 
   B_3=c_{l\sigma}^{\dag}c_{\bar l \sigma}$).

4) The contribution of a tunneling line of reservoir $r$ is 
$\gamma_{rl\sigma}^{\pm}(\omega_l)=\Gamma_{r}^l/2\pi~f^{\pm}
(\omega_l - \mu_r)$, taking the plus-sign
if the line is going backward with 
respect to the closed time path, and the minus-sign
if it is going forward. The same way the contribution of a bosonic line
is given by $b^{\pm}(\omega_l)= \alpha_{\rm ph}(\omega_l)/2\pi~n_{\rm b}^{\pm}(\omega_l)$.
%Here, $f(x)=f^{+}(x)=1-f^{-}(x),~n_{\rm b}(x)=n_{\rm b}^{+}(x)=n_{\rm b}^{-}(-x)$ are the
%Fermi and Bose functions.

5) There is an overall prefactor $(-i)(-1)^c$, where $c$ is the 
total number of vertices on the backward propagator plus the number of 
crossings of tunneling lines (no bosonic lines) plus the number of vertices 
connecting the state $d$ with $\uparrow$.

6) Integrate over the energies $\omega_l$ of the tunneling and boson lines and sum over 
all reservoir and spin indices. \\

\section{Results}
 
In the following we discuss current and shot noise for the model of
Eqs.~(1)-(4) with two single-particle levels ($M=2 \rightarrow l=1,2$) 
in first order perturbation theory 
in the tunnel couplings $\Gamma^r$ and the coupling $\alpha_{\rm ph}$ to the bosonic bath.
%A similar model (without Bose coupling and relaxation) was studied in Ref.~\onlinecite{nazarov}.
We express the different coupling parameters 
$\Gamma^r_l, \alpha_{\rm ph}$ in units of a scale $\Gamma$ that has the same 
order of magnitude as the largest of the tunnel couplings $\Gamma^r_l$.
(In the case of equal tunnel couplings, the natural choice is
$ \Gamma^{\rm R}_1 = \Gamma^{\rm L}_1 = \Gamma^{\rm R}_1 =\Gamma^{\rm L}_2=\Gamma$.)
Our perturbation expansion is valid for temperatures larger than the tunnel
couplings.
Throughout this paper, we choose $k_{\rm B} T = 10 \Gamma$.
%We consider transport through two levels which additionally may
%absorb or emit photons. 
The molecule can acquire 16 possible states, as each level can be either
unoccupied, occupied with spin $\uparrow$ or $\downarrow$, or doubly occupied.
The system described is characterized by level energies $\epsilon_1$ 
and $\epsilon_2$,
the 'Hubbard' repulsion $U$ and the charging
energy $E_c$. 
Furthermore the electron and photon reservoirs have temperature $T$ (set as $T=0.05meV$) 
%and $T_{\rm b}$
and are connected to the molecule via the coupling parameters
$\Gamma^r_l$ and $\alpha_{\rm ph}$. 
%We choose the scale $\Gamma = 0.1 T$ to be within the regime of validity of
%our perturbation theory *** need to introduce $k_{\rm B}$ ***.
Transport is achieved by applying a bias
voltage $V_{\rm bias}$, which is dropped symmetrically at the  electrode-molecule tunnel 
junctions, meaning that the energies of the molecular states are independent of 
the applied voltage even if the couplings $\Gamma^r_l$ are not symmetric. 
The effects of asymmetric voltage drop and a possible gate voltage are 
straightforward to anticipate, but would only obscure the results presented below.\\

\begin{figure}[h]
\centerline{\includegraphics[width=6.5cm]{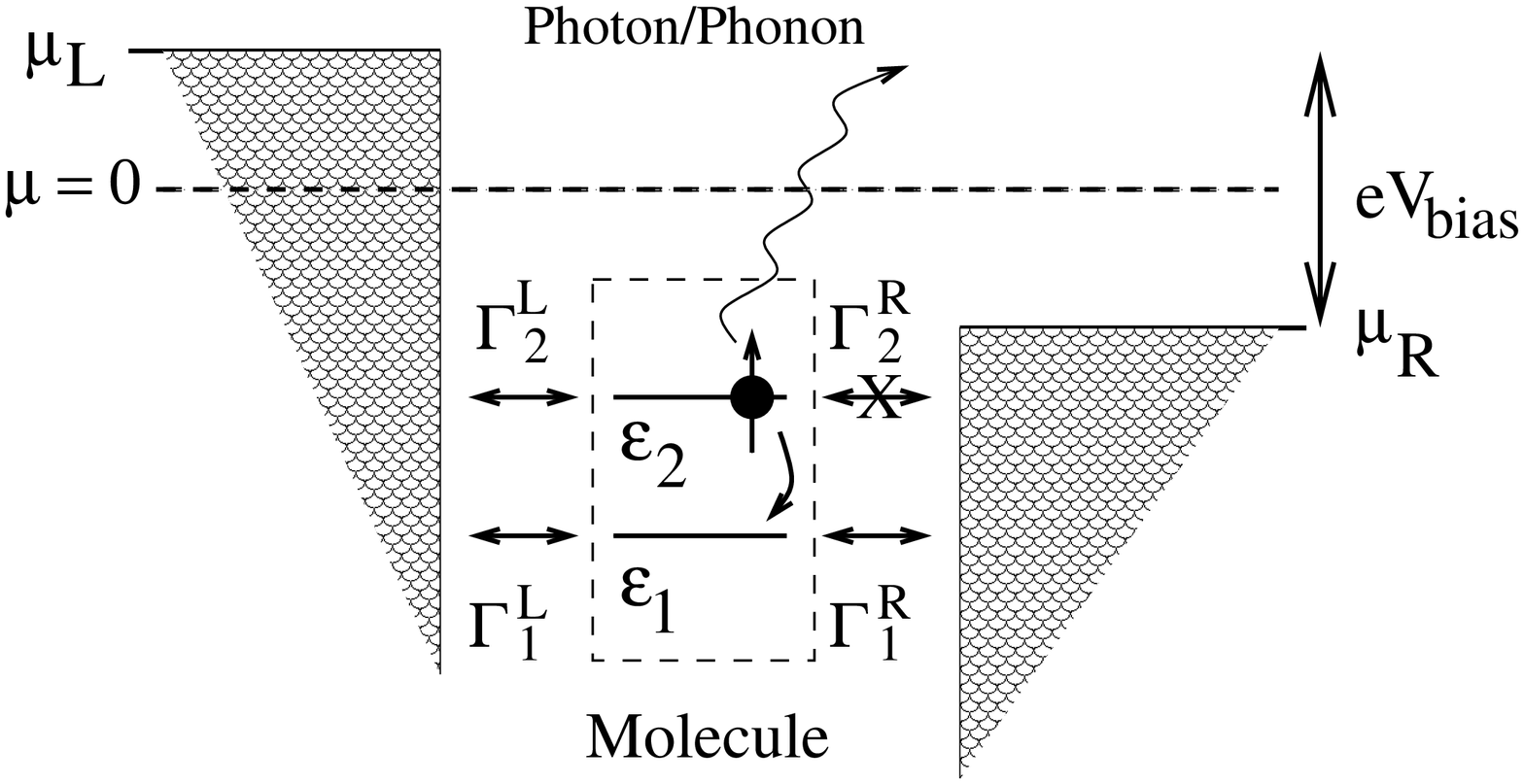}}
\caption{Sketch of a the couplings and processes in the considered model.}
\label{fig:tt_sketch}
\end{figure}

An illustration of the transport situation is shown in Fig. \ref{fig:tt_sketch}.
The Fano factor, which is given by the noise to current ratio, $F=S/2 e I$,
provides additional information about transport properties, not contained in 
the current-voltage characteristics alone. Therefore, we are interested in 
studying its dependence on different couplings to the electrodes, 
the strength of relaxation, the Coulomb charging energy, etc. 
in order to make predictions of the importance of 
those parameters for a given experiment.

\begin{figure}[h]
\centerline{\includegraphics[width=7.5cm, angle=270]{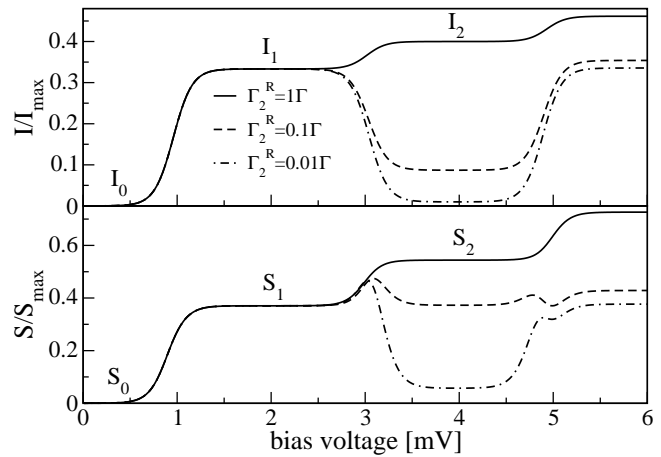}}
\caption{Current $I$ and shot noise $S$ vs. voltage for $k_{\rm B} T=0.05 {\rm meV}$, 
  $\epsilon_{1}=-0.5 {\rm meV}$, $\epsilon_{2}=0.5 {\rm meV}$,
  $U=1.5 {\rm meV}$, $E_c=1 {\rm meV}$, symmetric bias ($\mu_{\rm L}=-\mu_{\rm
  R}=eV/2$) and 
  $\Gamma^{\rm L}_1=\Gamma^{\rm L}_2=\Gamma^{\rm R}_1=\Gamma$. 
  The height of the plateaus labeled by $i=0,1,2$ are discussed in the text
  and depend on the choice of the coupling parameters. For suppressed coupling
  $\Gamma^{\rm R}_2$ current and shot noise break down leading to negative 
  differential conductance (NDC) at a threshold energy.
  The curves are normalized to $I_{\rm max}=(e/\hbar)2\Gamma$ and 
  $S_{\rm max}=(e^2/\hbar)2\Gamma$, respectively.
}
\label{fig:i-s-asym}
\end{figure}

We focus on the easiest case which exhibits NDC. 
The set of energy parameters 
$\{\epsilon_{1}=-0.5 {\rm meV},\epsilon_{2}=0.5 {\rm meV},U=1.5 {\rm meV},
E_c=1 {\rm meV} \}$\cite{footnote}  
describes a molecule that is uncharged at zero bias, as the energy to occupy
the first single particle level is $\epsilon_{1}+E_c=0.5 {\rm meV}$ (state $D_1$). Without coupling to the boson bath and 
%In the small bias regime (up
%to $7.5 mV$) transport through the molecule with states containing one electron
%can be modeled therefore for the underlying two level system. 
in contrast to the one level system discussed in Ref.~\onlinecite{thielmann} we find a
negative differential conductance (NDC) regime, see Figure~\ref{fig:i-s-asym},
%as can be seen in Figure~\ref{fig:i-s-asym} for the current $I$, 
in dependence on the different coupling strength between the molecular orbitals and the reservoirs,
as was previously discussed in Refs.~\onlinecite{hsw,bulka}. The shot noise behaves qualitatively similar,
the important quantitative details are discussed in the following. 

\begin{figure}[h]
\centerline{\includegraphics[width=6.8cm,angle=270]{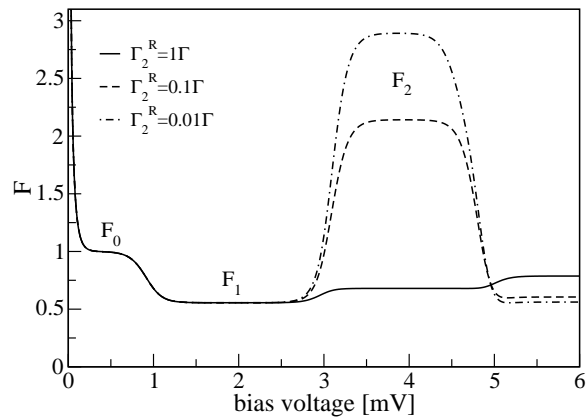}}
\caption{Fano factor $F$ vs. bias voltage for the same parameters as in 
  Fig.~\ref{fig:i-s-asym} and various coupling parameters $\Gamma^{\rm R}_2$.
The NDC effect results in a super-poissonian value for the Fano factor.}
\label{fig:fano-asym}
\end{figure}

If we chose equal tunnel coupling, $\Gamma^r_l=\Gamma$ (solid line), 
we find that current and shot noise $S$ increase, each time,
as a new transport "channel" (controlled by the excitation energies) opens. This leads
to plateaus, separated by thermally broadened steps. 
The first four plateaus are shown and discussed in the following.
At a bias voltage of $1{\rm mV}$, sequential transport through 
the state $D_1$ with one electron on the lower lying level becomes possible.
At $3{\rm mV}$, additionally transport through the $D_2$ state opens up, with the upper level being 
occupied with one electron. The different regions of interest are labeled by
$I_i,S_i$ with $i=0,1,2$.
For a bias voltage above $5{\rm mV}$, transport
channels with two or more electrons on the molecule open up. 
In the large-bias regime (not indicated in the plots) and for symmetric
coupling, the values $I_{\rm max}=(e/\hbar)2\Gamma$ and 
$S_{\rm max}=(e^2/\hbar)2\Gamma$ are reached. If now the coupling parameter
$\Gamma^{\rm R}_2$ is suppressed with respect to the other couplings, this leads 
to suppressed curves for the current and shot noise in region $2$, resulting
in NDC at the threshold of $3{\rm mV}$, when the state $D_2$ becomes relevant, see in Fig.~\ref{fig:i-s-asym} for $\Gamma^{\rm R}_2=0.1\Gamma$ and $0.01\Gamma$. 
The reason for the NDC is a combination of the Pauli principle, Coulomb
blockade and  suppressed coupling, as discussed in Ref.~\onlinecite{hsw,hettler_prl}.
In our case, an electron, entering the
upper molecular orbital from the left electrode, cannot leave the molecule,
if the coupling of this orbital to the right electrode is entirely suppressed. Transport through
the lower molecular orbital is also not possible, since the simultaneous
occupation of both orbitals is energetically forbidden in the considered bias regime. 
The electron gets stuck in the upper molecular orbital blocking other electrons
from tunneling through the molecule. Consequently, the current collapses.

Since in lowest-order perturbation theory in $\Gamma$ the plateau {\it heights} 
are given by the coupling parameters only, we find, 
that for $\Gamma^{\rm R}_2 < 2/3\Gamma$ NDC can be observed, whereas the shot
noise is suppressed below its lower bias  plateau only, if $\Gamma^{\rm R}_2<0.1\Gamma$. 
This difference can already give a rough idea about the coupling strength $\Gamma^{\rm R}_2$ for a given
set of current and noise measurements.
If the shot noise is sufficiently suppressed in the NDC region, a peak in the shot noise appears
around the resonance energy of the second level. 
This peak is due to temperature induced fluctuations that in certain 
situations enhance the shot noise over the surrounding  plateau values (where temperature fluctuations
are exponentially suppressed). As the resonance is approached from lower bias, within the range of
temperature broadening the noise "detects" the opening of the second transport channel and increases.
If the bias is beyond the resonance, the redistribution of occupation has taken place and the noise
is algebraically suppressed. The result is the observed peak with width of the temperature. However,
the peak height is only determined by the coupling parameters and is independent of the temperature.
The current never shows such a peak, as it decreases proportional to the loss of occupation of the first level,
the transport channel with "good" coupling.
% Peaks in shot noise

The effect of NDC on the Fano factor, which is given by $F=S/2eI$,
is shown in Figure~\ref{fig:fano-asym}.
At small bias, $eV \ll k_{B}T$, the noise is dominated by thermal noise,
described by the well known hyperbolic cotangent behavior which leads to a
divergence of the Fano factor.\cite{blanter,loss}
The plateau for bias voltages below $1 {\rm mV}$ (region $0$) corresponds to the 
Coulomb blockade regime, where transport is exponentially suppressed.
In the regions $1$ ($2$) transport through the state $D_1$ ($D_2$) is possible. 
The suppressed coupling strength $\Gamma^{\rm R}_2$
does not affect the plateau height $F_1$, whereas $F_2$ reaches values larger
then $1$, and up to $3$.\cite{bulka}  This "super-poissonian" noise ($F>1$) is predicted for 
$\Gamma^{\rm R}_2<0.44 \Gamma$. If the bias is larger then $5 {\rm mV}$ tunneling through states is allowed
where both orbitals are occupied simultaneously, i.e. the molecule can be doubly occupied.
The Fano factor is  sub-poissonian again ($F<1$) in this regime. 
Comparing the Figures~\ref{fig:i-s-asym} and \ref{fig:fano-asym} graphically allows
one to determine roughly the strength of the suppression of $\Gamma^{\rm R}_2$. 
In general, however, depending on the
underlying energy parameters (giving the ordering of a sequence of plateaus)
and the coupling parameters (giving their height) other values for the Fano
factor are possible. In the Coulomb blockade regime (region $0$), for example, 
there can  also be super-poissonian noise, if both energy levels are below
the equilibrium Fermi energy. Super-poissonian noise is also possible 
for a single Anderson level, if the spin degeneracy 
is lifted by a magnetic field (in the Coulomb blockade regime) or by 
ferromagnetic leads, see Refs. \onlinecite{belzig1,ferro_leads}.  
The energy and coupling parameters can be fully determined 
only by considering several transport
regimes, e.g. by application of a gate voltage.

It should be noted that the non-monotonic behavior of the Fano factor in regions $1$ and $2$ is entirely 
due to the second term of the noise expression Eq.~(\ref{S1}) that accounts for "propagation" (and transitions)
of molecule states between the two current vertices at different times.\cite{bulka} 
On the plateaus of regions $1$ and $2$ the first term of Eq.~(\ref{S1}) is always (i.e. for any coupling/relaxation parameter values) identical to the current times the electric charge $e$. Therefore, it contributes a term 1/2 to the Fano factor $F=S/2eI$.

\begin{figure}[h]
\centerline{\includegraphics[width=7.5cm,angle=270,]{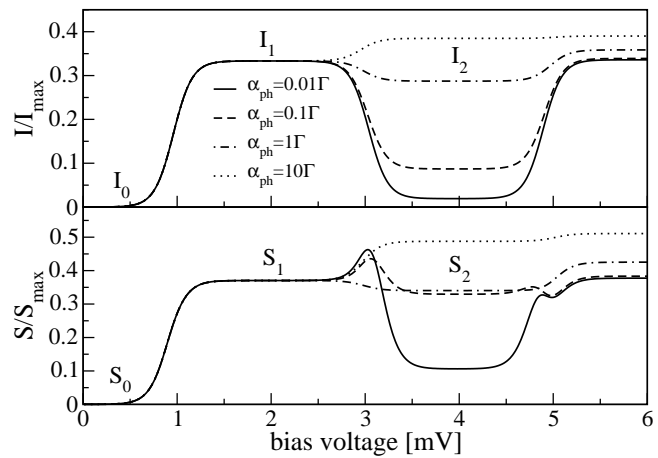}}
\caption{Current $I$ and shot noise $S$ vs. voltage for the same parameters as
in Fig.~\ref{fig:i-s-asym} but fixed coupling $\Gamma^{\rm R}_2=0.01\Gamma$.
Coupling to a bosonic bath allows for relaxation processes.  The coupling parameter
$\alpha_{\rm ph}$ which is varied relative to $\Gamma$. The NDC effect is
destroyed by strong relaxation.}
\label{fig:i-s-ph}
\end{figure}

\begin{figure}[h]
\centerline{\includegraphics[width=6.8cm,angle=270]{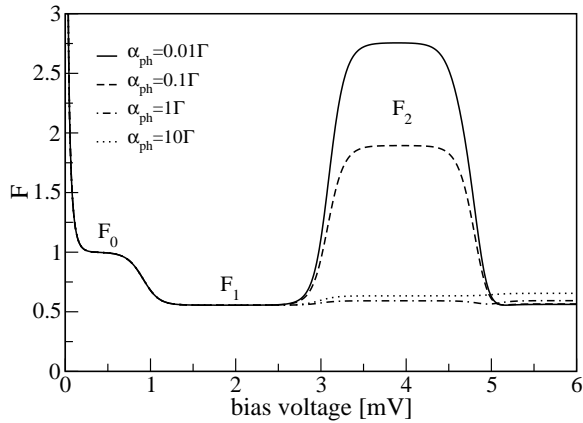}}
\caption{Fano factor vs. bias voltage for for the same parameters as in 
  Fig.~\ref{fig:i-s-ph} and various couplings to a bosonic bath $\alpha_{\rm ph}$.
The super-poissonian value of the Fano factor vanishes due to strong
  relaxation processes.}
\label{fig:fano-ph}
\end{figure}

\begin{figure}[t]
\centerline{\includegraphics[width=6.0cm,angle=270]{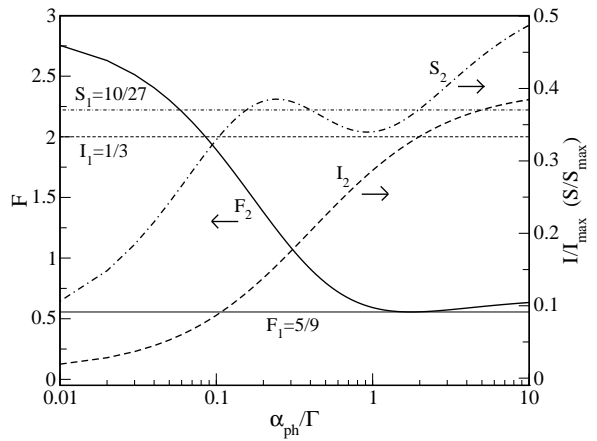}}
\caption{Fano factor (left axis), current and shot noise (right axis)
  vs. coupling to the bosonic bath for the same set of parameters as in 
  Figures~\ref{fig:i-s-ph},~\ref{fig:fano-ph}. 
  The values  $I_1,S_1,F_1$ of the first plateau do not depend on the bosonic 
  coupling parameter, whereas on second plateau (NDC region) 
  both the shot noise and the Fano factor show a non-monotonic dependence.}
\label{fig:isf-plateaus}
\end{figure}

Let us consider next the effect of relaxation processes on the current
and shot noise curves. In Figure~\ref{fig:i-s-ph} we keep the same set of
energy parameters as in Figure~\ref{fig:i-s-asym} and fix the coupling
strength $\Gamma^{\rm R}_2$ at $0.01\Gamma$ suppressed relatively to the 
other molecule-electrode couplings. 
Now a parameter
$\alpha_{\rm ph}$ describes the coupling of the molecule with a boson bath. 
A value of $\alpha_{\rm ph}=0.01\Gamma$ is below even the
relatively weak dipole coupling of photons to molecule states of small aromatic
molecules such as benzene.\cite{hettler_prl} 
For this small photon coupling (solid line) current and shot noise are
still reduced in region $2$ relative to the plateau heights $I_1=1/3$ and
$S_1=10/27$ in region $1$. If now $\alpha_{\rm ph}$ increases, we find that
both $I$ and $S$ also increase in the NDC region, at least initially. If the
value $\alpha_{\rm ph}=2\Gamma$ is exceeded, the NDC is gone
 (see also Figure~\ref{fig:isf-plateaus}, dashed line).
The behavior of the shot noise peak at the resonance energy is now further complicated 
by the effect of relaxation. The noise value at the resonance energy is non-monotonic,
i.e. it first decreases and then increases again with increasing relaxation. This is due
to redistribution of occupation by the relaxation processes in favor of the first level. 
 
For our chosen parameters, the value $\alpha_{\rm ph}=2\Gamma$ is larger than a reasonable
molecule-photon coupling. However, phonon (vibrational) couplings could easily be strong 
enough to achieve such fast relaxation. On the other hand, molecule vibrations have a discrete
spectrum, much different to the power law assumed in our calculations. Relaxation due to phonons
can be only effective, if the energies of a phonon and the electronic excitation
match within the smearing provided by temperature. This obviously depends on the details
of the molecule and can not be discussed within the model considered here.
The destruction of NDC by bosonic transition rates is easily explained. An
electron which formerly was stuck on the upper molecular orbital can now relax onto the
lower molecular orbital, from which tunneling to the right electrode
is possible via the coupling $\Gamma^{\rm R}_1$. 

For the Fano factor in Figure~\ref{fig:fano-ph} an increase of $\alpha_{\rm ph}$ leads to a decreasing value for the plateau $F_2$, which passes the poissonian value
$F=1$ at $\alpha_{\rm ph} \sim 0.34\Gamma$. Different to the current, however, the Fano
factor does not show monotonic behavior with increasing $\alpha_{\rm ph}$.
The dashed-dotted line corresponding to $\alpha_{\rm ph}=1\Gamma$
lies below the dotted one with $\alpha_{\rm ph}=10\Gamma$. The non-monotonic behavior
is even more pronounced for the shot noise. It has a maximum and a
minimum for $0.2 \sim \alpha_{\rm ph}/\Gamma \sim 1$ before increasing again at $\alpha_{\rm ph} > \Gamma$,
see Figure~\ref{fig:isf-plateaus}. The richness of the noise behavior in the NDC
regime might allow a detailed determination of coupling-parameter values.

Contrary to the (monotonic) $\alpha_{\rm ph}$-dependence of the current, which can be explained by 
a redistribution of occupation probability form the "blocking" upper level to the "conducting" lower level,
it is difficult to present a simple physical picture for the non-monotonic shot noise behavior in the NDC region.
As noted above, it is the second term of the noise
expression Eq.~(\ref{S1}) that is responsible for variation of the noise with coupling parameters.
For our set of parameters, the second term of Eq.~(\ref{S1}) has a peak at about  $\alpha_{\rm ph} \sim 0.16 \Gamma$
and a minimum at $\alpha_{\rm ph} \sim 1.66 \Gamma$, where it almost reaches zero. The increase of this term at small 
$\alpha_{\rm ph}$  is again explained by the lifting of the blockade, i.e. the redistribution of 
occupation probability. The decrease after the maximum is the result of a (near) cancellation of contributions 
from the different states participating in transport. Some of the state contributions are negative and counter the positive contributions that produce the maximum. Such a non-monotonicity  with coupling parameters is 
only possible (in first order perturbation theory) for a reducible observable like the shot noise, where aside of the
stationary occupation probabilities also the "molecule propagator" (in the form of {\bf P} of Eq. \ref{def_P})  plays an important role.

Since in lowest-order perturbation theory
temperature only leads to a thermal broadening of the steps, the plateau
heights in the different transport regimes are given by the coupling
parameters, both the tunnel coupling as well as the relaxation strength. 
However, note that the actual relaxation {\it rate} depends also on the position of the energy levels
via the boson density of states. This will complicate  matters in the general case
with many levels, which are not equidistant from each other. 
In our case with two levels, we can extract analytical
expressions for the plateau values current, noise and Fano factor within the
low bias transport
regimes as indicated in the Figures~\ref{fig:i-s-asym} to \ref{fig:fano-ph}.
We find for the plateau of the NDC-region $2$ ($S_2=2I_2F_2$)
\begin{equation}
I_2 ={\Gamma^{\rm R}_1(\alpha_{\rm ph}+\Gamma^{\rm R}_2)(\Gamma^{\rm
    L}_1+\Gamma^{\rm L}_2)/\Gamma
\over 
2\Gamma^{\rm L}_2(\alpha_{\rm ph}+\Gamma^{\rm R}_1)+
(2\Gamma^{\rm L}_1+\Gamma^{\rm R}_1)(\alpha_{\rm ph}+\Gamma^{\rm R}_2)}
\end{equation}
for the current and
\begin{widetext}
\begin{equation}
F_2={\alpha_{\rm ph}(\alpha_{\rm ph}+2\Gamma^{\rm R}_2)[(\Gamma^{\rm R}_1)^2+
4(\Gamma^{\rm L}_1+\Gamma^{\rm L}_2)^2]
+ [8\Gamma^{\rm L}_1\Gamma^{\rm L}_2(\Gamma^{\rm R}_1-\Gamma^{\rm R}_2)^2+
4(\Gamma^{\rm L}_1\Gamma^{\rm R}_2+\Gamma^{\rm L}_2\Gamma^{\rm
  R}_1)^2+(\Gamma^{\rm R}_1\Gamma^{\rm R}_2)^2]
\over 
[ 2\Gamma^{\rm L}_2 (\alpha_{\rm ph}+\Gamma^{\rm R}_1)  +
(2\Gamma^{\rm L}_1+\Gamma^{\rm R}_1)(\alpha_{\rm ph}+\Gamma^{\rm R}_2)]^2}.
\end{equation}
\end{widetext}
for the Fano factor. Since only bosonic transition between singly occupied levels $1$ and $2$ are 
possible in this bias region, the above expressions include only one bosonic rate 
$\alpha_{\rm ph}(\Delta E = \epsilon_2 - \epsilon_1)$. Since the temperature is
much smaller than $\Delta E$, only relaxation processes matter for the plateau values.
For completeness, we also give the expressions for the
transport regime $1$ (transport through the lower level only).
They can be found from the above  
by setting the couplings $\Gamma^{\rm L}_2$ and $\alpha_{\rm ph}$ equal to $0$. Then electrons can never enter
the upper level at positive bias, leading to an effective one level system with 
the result\cite{thielmann}
\begin{equation}
I_1={2\Gamma^{\rm R}_1\Gamma^{\rm L}_1
\over 
(2\Gamma^{\rm L}_1+\Gamma^{\rm R}_1)}{1 \over 2\Gamma}\; ; \; \; 
F_1={4{(\Gamma^{\rm L}_1)}^2+{(\Gamma^{\rm R}_1)}^2
\over 
(2\Gamma^{\rm L}_1+\Gamma^{\rm R}_1)^2}.
\end{equation}

\begin{figure}[b]
\centerline{\includegraphics[width=6.8cm,angle=0]{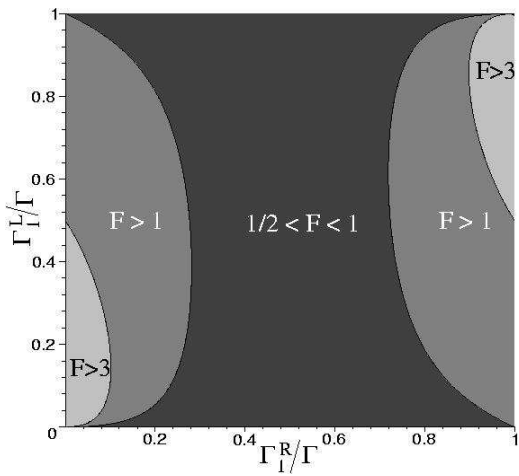}}
\caption{Contourplot of the Fano factor (plateau $F_2$) with the choice
  $\Gamma^{\rm L,R}_1=\Gamma-\Gamma^{\rm L,R}_2$ and $\alpha_{\rm ph}=0$. The totally symmetric
  situation is given for $\Gamma^{\rm L}_1=\Gamma^{\rm R}_1=0.5 \Gamma$.
  The Fano factor can become arbitrarily large, if the system is sufficiently asymmetric.
}
\label{fig:fano-2D}
\end{figure}

The derivation of analytical expressions in the low bias regime allows us
a quick study of current, noise and Fano factor for arbitrary coupling situations.
For the special situation where $\Gamma^{\rm L,R}_1=\Gamma-\Gamma^{\rm L,R}_2$ and $\alpha_{\rm ph}=0$ the Fano
factor $F_2$ is presented in a contourplot (see Figure~\ref{fig:fano-2D}). 
This choice allows the coupling parameters to the left and right reservoir to vary (independently) 
between 0 and $\Gamma$, while 
having the sum of the couplings to each reservoir fixed. Although
not all of the possible coupling situations can be visualized this way, the 
following features which can be extracted from this plot are valid in general:
A super-poissonian Fano factor $F>1$ can only be found, if $\Gamma^{\rm
  R}_1\neq\Gamma^{\rm R}_2$ and additionally  $\Gamma^{\rm
  L}_1\neq0,\Gamma^{\rm L}_2\neq0$.
Furthermore a Fano factor $F>3$ is possible only if $\Gamma^{\rm
  L}_1\neq\Gamma^{\rm L}_2$
besides the above conditions. 
In the absence of relaxation processes ($\alpha_{\rm ph}=0$) we can also find a
point symmetry of $F_2$.
%if couplings $\Gamma^r_1 \leftrightarrow \Gamma^r_2$
%are exchanged, which is obvious, since we have two independent levels. 
This symmetry is broken if $\alpha_{\rm ph} \neq 0$, as adsorption and emission
rates of bosons differ due to the boson occupation factors.

%$F_2(\alpha_{\rm ph})=const.$ if $\Gamma^R_1=\Gamma^R_2$ or $\Gamma^L_2=0$
%($F_2:[1/2;1]$)\\

The special case with the settings
$\Gamma^{\rm L}_l \rightarrow \Gamma^{\rm L}_l/2$
describes spinless transport through a two level-system in the absence of relaxation
processes. This situation was discussed in Ref.~\onlinecite{kiesslich2} where values of the
Fano factor between $1/2$ and $3$ were found in the case of equal couplings
$\Gamma^{\rm L}_1=\Gamma^{\rm L}_2$. In the case  $\Gamma^{\rm L}_1\neq\Gamma^{\rm L}_2$
%of suppressed couplings of
%the upper level to the reservoirs compared to the lower level-electrode couplings
we can even find Fano factors with values much larger than $3$, as the shot
noise is strongly enhanced compared to a current that is still sizable itself (not exponentially suppressed), see Figure~\ref{fig:fano-2D}.
This again only happens for a special set of coupling parameters, thus
allowing a detailed analysis of the coupling parameters, if such high values for the Fano factor 
were observed in experiment.

Besides the super-poissonian noise with Fano factors  $F > 1$ due to
positive correlations and values between $1/2< F < 1$ in the sub-poissonian
regime, we can also find values of coupling parameters in which the Fano
factor drops to values below $1/2$. This behavior, however, can only be observed in
the presence of relaxations, when the coupling strength
$\Gamma^{\rm L}_1$ and $\Gamma^{\rm R}_2$ are suppressed relative to the other
tunnel couplings. If the above couplings are vanishing, there is only one path for the
electrons to tunnel through the molecule, namely from the left electrode
to the upper molecular orbital, then via relaxation onto the lower molecular orbital
until finally the electrons leave the molecule by tunneling to the right electrode. By choosing specifically
$\Gamma^{\rm L}_1=\Gamma^{\rm R}_2=0$ and 
$\Gamma^{\rm R}_1=2\Gamma^{\rm L}_2=\alpha_{\rm ph}$ the value of the Fano factor 
can be minimized and is found to be $1/3$. The probabilities to find an
unoccupied molecule or an occupied molecule with one electron in the lower level doublet or in upper level doublet
are all equal in this case ($P_0=P_{D_1}=P_{D_2}=1/3$). 
This special situation reminds of a system, where a chain of 
quantum dots are coupled in series, having interdot tunnel couplings of the
same size than the couplings of the chain ends to the leads. 
For an infinite chain of such dots (effectively a one-dimensional wire) 
the Fano factor also reaches $1/3$.\cite{beenakker,nagaev}

In summary, we have discussed the interplay of Coulomb interactions, level-dependent 
coupling and relaxation in a model suitable for quantum dots, molecules and short nanotubes.
We find super-poissonian shot noise in a bias region where the electronic current is suppressed 
due to blocking effects. Relaxation due to bosonic excitations 
has a strong impact on both current and shot noise in this 
region, for example, the shot noise behaves non-monotonically as a
function of the coupling strength to the bosonic bath.
The Fano factor (noise to current ratio) can become arbitrarily large in the blocking regime. 
In another special set of couplings the Fano factor can be reduced to 1/3, resembling
that of a one-dimensional wire.

{\em Acknowledgments.}
We enjoyed interesting and helpful discussions with G. Kiesslich
as well as financial support by the DFG via the Center for 
Functional Nanostructures and the Emmy-Noether program.


\begin{thebibliography}{99}
\bibitem{footnote}The chosen energy values are appropriate for a quantum dot or short nanotube.
For molecules, one would expect energies that are a factor of 100-1000 larger (roughly, $meV \rightarrow eV$).
\end{thebibliography}
\end{document}